\documentclass[portuguese,
  manuscript=article,  
  layout=publish,
  year=2024,
  volume=,
]{jmfis}

\usepackage[brazilian]{babel} 
\usepackage{amsmath}
\usepackage{dsfont}
\usepackage{bm}
\usepackage{cite}
\usepackage{graphicx}
\hypersetup{pdfborder = 0 0 0}

\usepackage{xcolor}

\definecolor{lightblue}{HTML}{0000FF}

\definecolor{indigo}{HTML}{4B0082}

\doi{{ {\color{blue}\href{}{}}}}

\received {}
\accepted {19/11/2024}
\published{}

\newcommand{\orcidlink}[1]{\href{https://orcid.org/#1}{\includegraphics[width=10pt]{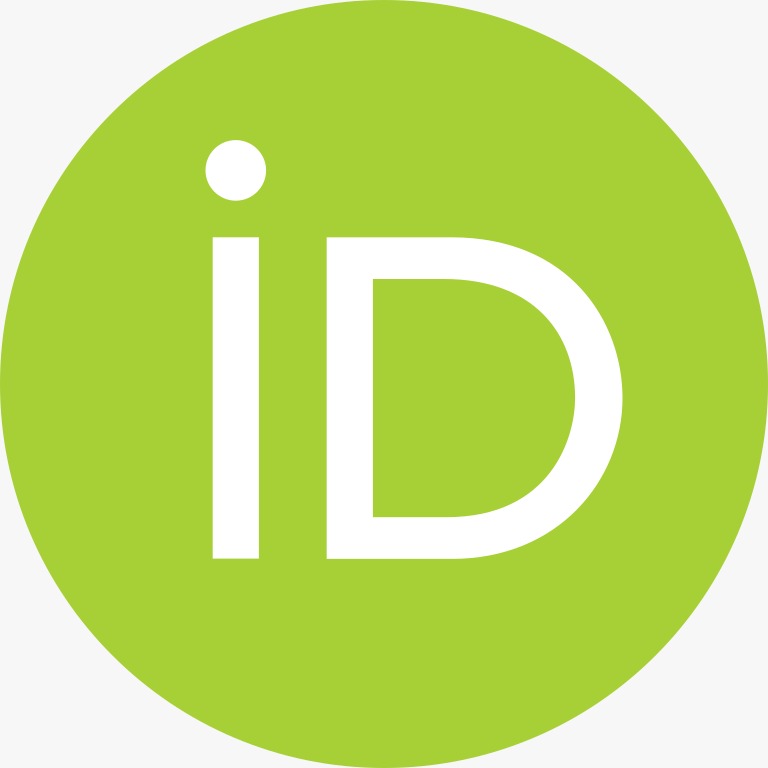}}}

\title{Uma proposta para o uso de RPG no Ensino de Física: A Vingança de Newton}

\author{Maria Rita Vasconcelos Brandão Souza  \orcidlink{0000-0000-0000-0000}}
\email{brandaomariar@gmail.com}
\affiliation{Laboratório de Engenharia de Processos, Faculdade de Engenharia de Alimentos da Universidade Estadual de Campinas, Rua Monteiro Lobato, 80, Cidade Universitária, 13083-862, Campinas, SP, Brasil}

\author{Luís Henrique de Freitas \orcidlink{0000-0000-0000-0000}}
\email{luis.h.freitas@ufv.br}
\affiliation{Universidade Federal de Viçosa
Rua José da Cruz Reis, 158, Centro, 36570-071, MG, Brasil}

\author{Glaucia de Souza Silva  \orcidlink{0000-0000-0000-0000}}
\email{glauciasos@gmail.com}
\affiliation{Instituicao atual: Colégio Luterano, Av. Herman Teles Ribeiro, 162, Vila Romanopolis, 08529-100, Ferraz de Vasconcelos, SP, Brasil}

\author{Felipe Xavier de Carvalho \orcidlink{0000-0000-0000-0000}}
\email{fxclsmg@gmail.com}
\affiliation{Instituto Federal de Educação, Ciência e Tecnologia do Rio Grande do Norte - IFRN - Campus Natal Central CNAT, Rua Historiador Francisco Fausto de Souza, 337, 59082-260, Campim Macio, RN, Natal}

\author{Leonardo Antônio M. Souza \orcidlink{0000-0001-8374-984X}}
\email{leonardoamsouza@ufv.br}
\affiliation{Universidade Federal de Viçosa - Campus Florestal, 35690-000, Florestal, MG, Brasil}

\palavrachave{Metodologias alternativas de Ensino; Processo Ensino-Aprendizado; Gamificação no Ensino}

\begin{document}
\begin{resumo}
Este trabalho apresenta uma proposta para a exploração do uso de Role Playing Games (RPG) como metodologia ativa no ensino de Física Moderna, com foco em um jogo chamado A Vingança de Newton. O jogo foi desenvolvido com o objetivo de engajar os estudantes em processos de aprendizado colaborativo e investigativo, utilizando elementos de gamificação para aumentar a motivação e o envolvimento. Funda-\newline mentado nas teorias construtivistas de Piaget e Vygotsky, o RPG estimula o desenvolvimento cognitivo e social, ao colocar os estudantes no papel de personagens históricos da ciência. Através de desafios físicos contextualizados, como a compreensão do Efeito Fotoelétrico, os participantes constroem conhecimento de forma participativa. Este estudo apresenta dados preliminares de aprendizagem, obtidos por meio de pré e pós-testes, além de avaliar a percepção dos estudantes quanto ao uso de jogos educativos no ensino de ciências. Os resultados indicam que o uso do RPG pode ser uma ferramenta eficaz para o ensino de Física Moderna, promovendo um maior engajamento e compreensão dos conceitos científicos.
\end{resumo}

\vspace{0.5cm}

\begin{abstract}
This work explores the use of Role Playing Games (RPG) as an active methodology in teaching Modern Physics, focusing on a game called Newton's Revenge. The game was developed with the aim of engaging students in collaborative and investigative learning processes, using gamification elements to increase motivation and involvement. Based on the constructivist theories of Piaget and Vygotsky, the RPG stimu-\newline lates cognitive and social development by placing students in the roles of historical science figures. Through contextualized physical challenges, such as understanding the Photoelectric Effect, participants actively construct knowledge. This study presents preliminary learning data obtained through pre- and post-tests, as well as evaluates students' perceptions of using educational games in science education. The results indicate that the use of RPG can be an effective tool for teaching Modern Physics, promoting greater engagement and understanding of scientific concepts.
\end{abstract}

\section{Introdução}

Este artigo visa detalhar e contextualizar parte da apresentação de um dos autores na 20$^a$ Escola Matogrossense de Física, realizada em Outubro de 2024.

Aproximadamente no século IV antes da Era Comum viveu Aristóteles, filósofo e polímata grego. Dentre suas várias abordagens, Aristóteles desenvolveu o que ficou conhecido como as Quatro Causas de Aristóteles \cite{1}. Um dos pensamentos que, aparentemente, permeava a mente de Aristóteles era: Como saber que uma determinada ``coisa'' é de fato esta ``coisa''? Aristóteles argumentava que podemos pensar em alguns ``porquês'' para responder o que de fato é alguma coisa, que ele chamou de \emph{Causas}. Responder às Causas de uma ``coisa'' dava o significado real daquela ``coisa''. De acordo com Aristóteles existem Quatro Causas para caracterizar o conhecimento sobre uma certa ``coisa'', são elas: 1) Causa Material: do que é feita a ``coisa''. 2) Causa Formal: o formato da ``coisa''. 3) Causa Eficiente (ou Agente): quem ou o que fez a ``coisa''. 4) Causa Final: o propósito da ``coisa''. Por exemplo, uma mesa: para tentar entender as Quatro Causas Aristotélicas, podemos pensar que uma mesa pode possuir as seguintes Causas: 1) Causa Material: do que é feita a mesa, por exemplo madeira. 2) Causa Formal: o design da mesa. 3) Causa Eficiente: a pessoa que trabalhou a madeira e executou o design, um(a) marceneiro(a). 4) Causa Final: o objetivo da mesa, por exemplo ser uma mesa de jantar. Ao responder a estas Quatro Causas podemos ter conhecimento completo, segundo Aristóteles, acerca da mesa em questão, pois dessa forma saberemos que não é uma mesa de sinuca, mas uma mesa de jantar. Também saberemos que é uma mesa feita de madeira por algum artesão, e também qual o design da mesa. Sendo assim, pelo menos em princípio, podemos encontrar as Quatro Causas de Aristóteles para alguma ``coisa'' em questão, com o objetivo final de ter conhecimento sobre esta determinada ``coisa'' e, de repente, para podermos estudar uma destas Causas para realizar alguma mudança na ``coisa'' que estamos interessados(as).

Podemos por exemplo aplicar o conceito das Quatro Causas Aristotélicas para o Processo Ensino-Aprendizagem. Por exemplo, podemos conjecturar que suas Quatro Causas Aristotélicas são: 1) Causa Material: os(as) estudantes da turma. 2) Causa Eficiente: os agentes causadores do processo ensino-aprendizagem; apenas para fins deste trabalho, podemos enxergar esta causa como Estudantes e Professor(a), porém isto não seria talvez suficiente, pois esta causa envolve também toda a estrutura e gestão escolar como agentes neste processo. 3) Causa Formal: a sala de aula, virtual ou presencial. 4) Causa Final: qual tipo de conhecimento o(a) estudante deva possuir ao final da disciplina ou curso.

Quando falamos das chamadas ``Metodologias Tradicionais de Ensino'', em geral pensamos nas Quatro Causas da seguinte maneira (neste ponto, estamos generalizando a abordagem, para trabalhar a esta Introdução): 1) Causa Material: os(as) estudantes da turma. 2) Causa Eficiente: em geral o(a) docente é agente de ``transformação'', ou é um agente de transferência de conhecimento. O(a) docente é a Causa Eficiente central. 3) Causa Formal: a sala de aula, virtual ou presencial, com estudantes enfileirados (presencial), com salas em possível formato de arena, em silêncio, com poucas perguntas, anotando o conhecimento transferido pelo(a) docente. A sala de aula é a Causa Formal, e o(a) docente é ponto focal neste ambiente. 4) Causa Final: o(a) estudante adquire o conhecimento (total ou parcial), e é avaliado(a) para fins de aprovação, em geral a avaliação é individual e realizada de maneira estatística, com grande número de estudantes em cada avaliação/turma. Desta forma, podemos pensar as ``Metodologias Tradicionais de Ensino'' como um Ensino Centrado no(a) Docente, ou na Figura do(a) Professor(a).

Em contrapartida, tem-se estudado bastante as chamadas ``Metodologias Alternativas de Ensino'', ou ``Metodologias `Ativas' ''. Neste tipo de abordagem metodológica, em geral a figura do(a) Estudante é vista como central no Processo Ensino-Aprendizagem, como pode ser observado nas Quatro Causas Aristotélicas que a equipe deste trabalho a caracteriza: 1) Causa Material: continua sendo os(as) estudantes da turma. 2) Causa Eficiente: o(a) estudante é o próprio objeto de ``transformação'', ou é seu próprio agente de aquisição de conhecimento. O(a) estudante é a Causa Eficiente central, e o(a) docente é um guia, um(a) Tutor(a) para orientar o(a) estudante em sua busca por conhecimento. 3) Causa Formal: a sala de aula, virtual ou presencial, com estudantes organizados em subgrupos, dispostos em pequenos círculos (se presencial) ou sub-salas virtuais (se remoto), em geral a turma discute e debate os problemas de maneira entusiasmada (ou isto é o que se espera), perguntas surgem a todo momento. Os subgrupos da sala de aula, formado por estudantes, tornam-se a Causa Formal. 4) Causa Final: o aprendizado pode ser medido também por avaliações, porém o processo de obtenção do conhecimento em subgrupos e com Metodologias diferenciadas torna-se parte integrante da avaliação. Além disto, habilidades sociais dos(as) estudantes acabam sendo desenvolvidas por cada tipo de Metodologia que pode ser trabalhado. Comparando as Quatro Causas apresentadas acima para cada tipo de abordagem Metodológica, Tradicional ou Alternativa, notamos que a primeira é claramente centrada na figura do(a) Docente, como dito acima, e a segunda se torna um    Ensino Centrado no(a) Estudante. Evitaremos rotular certo tipo de Metodologia como ``Passiva'' ou ``Ativa'' como é comum na Literatura do assunto, pois entendemos que seja improvável um tipo de aquisição de conhecimento completamente ``Passivo'' por parte dos(as) estudantes. Chamaremos de Ensino Centrado no(a) Estudante os tipos de Metodologia que apresentaremos.

Dada esta Introdução, é possível vislumbrar que nossa intenção é salientar a importância de estudarmos e executarmos um tipo de Ensino Centrado no(a) Estudante, de modo a trabalhar a Causa Eficiente e Final com estas abordagens alternativas de Ensino. Nosso objetivo é justamente utilizar um tipo de Metodologia de Ensino Centrada no(a) Estudante (MECE), no caso a Gamificação e em especial o uso de RPG (Role Playing Games), para o Ensino de Física Moderna. No decorrer deste texto, daremos uma definição um pouco mais precisa do que estamos chamando de Ensino Centrado no(a) Estudante. Apresentaremos brevemente alguns tipos de MECE, incluindo algumas Metodologias já trabalhadas no Campus UFV-Florestal com algum sucesso. Em seguida, discutiremos a ideia por trás da Gamificação do Ensino, seus prós e alguns contras. Por fim, discutiremos rapidamente sobre o que é um RPG, e detalharemos o jogo que foi desenvolvido por nosso grupo.

\section{Referencial Teórico Básico}

Como será mencionado em uma seção futura deste trabalho, o RPG é um tipo de jogo colaborativo, e quando usado no contexto do Ensino pode trazer benefícios \cite{6,7,8,9,10,11,12,13}. Nossa abordagem neste trabalho é analisar uma forma lúdica, a partir de um jogo de RPG, para realizar algo próximo de um Ensino por Investigação \cite{18}, sendo que podemos mencionar Piaget \cite{19} e Vygotsky \cite{20}. Este trabalho é embasado, básica e fundamentalmente, em Teorias Construtivistas, principalmente Ensino por Investigação. Segundo Piaget \cite{19}, apesar de ele propriamente não ter construído exatamente uma Teoria de Aprendizagem mas sim uma Teoria de Desenvolvimento Mental/Cognitivo, para a construção do conhecimento é importante a introdução de uma situação-problema, que faz com o(a) estudante saia de sua posição de conforto de modo a raciocinar e construir seu conhecimento. Ao construí-lo, o(a) estudante terá condições de socializar, expor suas ideias com seus colegas, para assim chegarem às suas conclusões. Ainda citando Piaget, ``é necessário estabelecer entre as crianças, sobretudo entre os adolescentes, relações sociais, apelar para a sua atividade e para a sua responsabilidade'' \cite{19}.

A última frase do parágrafo anterior, de Piaget, pode ser relacionada com parte da Teoria socio-construtivista, desenvolvida por Vygotsky, que enxergava o desenvolvimento de uma pessoa sob uma perspectiva sócio-cultural (ou sócio-histórica), isto é, um(a) indivíduo(a) se estabelece em sua interação com o meio que está inserido. Portanto, as relações sociais em sala de aula são bem aceitas como fundamentais para o desenvolvimento do(a) estudante \cite{20}. Além disto, para Vygotsky é importante também o papel de uma mediação, de instrumento que mediam a interação da pessoa com o mundo que a circunda. Estes instrumentos podem ser técnicos ou sistemas de signos. Podemos entender um Jogo, em especial um RPG, como um instrumento mediador entre a pessoa que joga e o próprio mundo do jogo. Em nosso trabalho, o Universo criado por nós dentro do jogo de RPG acaba por fornecer à pessoa que joga métodos de interação com um fenômeno de Física Moderna que, em geral, não faz tanto parte do dia-a-dia da pessoa.

Podemos pensar em uma união dos pensamentos acima (obviamente respeitando as diferenças entre cada abordagem dos autores citados), dentro de um processo de Gamificação, isto é, do uso de jogos ou da mecânica de jogos no processo Ensino-Aprendizado, e em especial usando um jogo de RPG, quando bem estudado e executado, da seguinte forma: esta Metodologia faz uso de conhecimento anterior do(a) Estudante; auxilia nas atividades em grupo; na socialização do conhecimento; se usarmos as Salas em progressão como propomos em nossa primeira aventura completa, permitem que o(a) Estudante adquira conhecimento através de um ensino tipo investigativo. São vários os trabalhos que utilizam processos de Gamificação no Ensino (por exemplo \cite{6,7,8,9,10,11,12,13}).

\section{Metodologias Alternativas de Ensino}

Uma Metodologia de Ensino pode ser vista como uma diretriz que guiará o processo de Ensino e Aprendizagem, contendo estratégias, abordagens metodológicas, técnicas de Ensino, etc. Por exemplo, o Ensino tradicional, o chamado ``cuspe e giz'', é um tipo de Metodologia de Ensino, sendo definida por nós como Metodologia Tradicional de Ensino como explicado na Introdução deste trabalho (ver também \cite{2}). Definimos uma Metodologia de Ensino Centrada no(a) Estudante (MECE) como uma estratégia de Ensino centrada na participação efetiva de Estudantes, em geral divididos em subgrupos dentro de uma sala, na construção do processo de aprendizagem \cite{2}. O Docente utilizando alguma MECE tem o importante papel de tutoria, de orientador(a), um guia para que o(a) Estudante construa o conhecimento.

Não é a função deste trabalho explicar em detalhes as MECE mais utilizadas, apenas as citaremos por motivo de completeza. São vários os tipos de Metodologias Alternativas de Ensino disponíveis ``no mercado''\footnote{Inclusive deixamos claro que os autores entendem a questão ``mercadológica'' das várias alternativas metodológicas. Não estamos aqui querendo dizer que \emph{nossa} metodologia é melhor ou pior, mas que serve como alternativa para o processo Ensino-Aprendizado.} atualmente, sendo que são adaptáveis para a realidade de cada pessoa ou região. Podemos citar:
\begin{itemize}
\item Think-Pair-Share (pensar, compartilhar, socializar);
\item Instrução por Pares (Peer Instruction) \cite{3};
\item Sala de aula invertida;
\item Problem-based Learning (Aprendizagem baseada em problemas);
\item Aprendizagem baseada em projetos \cite{4};
\item Gamificação \cite{5}: esta abordagem será o foco de estudo neste trabalho. Daremos mais detalhes sobre a Gamificação na próxima seção.
\end{itemize}

\section{Gamificação}

A ideia central da Gamificação no Ensino é, como foi dito acima, usar elementos de jogos para trabalhar conteúdos em sala de aula, aumentando o engajamento e a motivação de estudantes \cite{5}. São várias as aplicações da Gamificação no Ensino, ver por exemplo as referências \cite{5}. Portanto, o propósito do uso de Metodologias de Ensino baseadas em Gamificação é usar jogos, ou mecânicas/dinâmicas advindas de jogos, ou aulas roteirizadas com linguagem de jogos.

Pode-se usar jogos de computador, jogos de cartas/tabuleiros, ou mesmo elementos de jogos para tentar abordar algum tema. Assim, características intrínsecas de jogos são utilizadas no processo de Gamificação, como \cite{5}: desafios explícitos a serem cumpridos em cada etapa; recompensas para seu personagem após cada etapa ser cumprida; um certo tipo de ``progressão'' do(a) Estudantes (como a passagem de ``fases''); motivação para completar o objetivo do jogo; para o caso específico do RPG, como será visto abaixo, a cooperação entre estudantes e docente se torna primordial para que se complete a tarefa. Sendo assim, é possível o treinamento de habilidades sociais dentro do grupo; dado que o ambiente do jogo é ficcional, e também pode-se criar planejamento da pessoa ou do grupo para tomar certos ``riscos''; há claramente e de imediato um retorno sobre erros e sobre a performance de cada pessoa; dependendo do tipo de jogo trabalhado, podemos deixar que estudantes ``brinquem'' com ideias.

Um retorno sobre os erros promove a persistência dos(as) Estudantes na tarefa a ser cumprida; requer um certo foco da pessoa para acompanhar o desafio em questão. Sendo assim, a proposta de uma Gamificação no Ensino é: \textbf{abordar certo conteúdo ou tema, utilizando aspectos de Jogos (RPG, tabuleiro, video-game, cartas, etc), para que as pessoas se sintam motivadas, engajadas a aprender o tema. A partida deve ser construída a dar uma ideia de progressão, do jogo, mas principalmente do conhecimento adquirido pelo grupo}.

\subsection{RPG (Role Playing Game)}

O RPG (Role Playing Game) é um jogo de interpretação (Role Playing, Interpretação de Personagem) \cite{6,7,8,9,10,11,12,13}. É um jogo de ``faz de conta'', de ficção, que pode acontecer em qualquer Universo ou época, o que dá muita liberdade criativa para quem está jogando e também para quem está narrando o jogo. Em geral os participantes criam e assumem o papel de seu personagem durante toda a história que está sendo interpretada. Para jogar um RPG o grupo de pessoas se reúne para narrar uma história, em geral estes grupos são chamados de ``Mesa''.

As regras de um RPG estão contidas em seu Sistema de Jogo, sendo que existem vários disponíveis, desde os mais complexos como Dungeons\&Dragons; até os mais simples como Guaxinins\&Gambiarras (G\&G) \cite{14}. Para entender as regras básicas de qualquer RPG, notamos que cada personagem possui um ou mais Atributos, características que o definem, e estes Atributos são em geral finitos. Com os valores dos Atributos, a Ação que a personagem for efetuar pode resultar em um Acerto ou em um Erro, sendo que isto é definido através da rolagem de algum tipo de dado. Através do valor do Atributo e da rolagem do dado existe uma probabilidade da Ação resultar em um acerto ou em um erro.

Além do Sistema de Jogo, duas figuras são fundamentais para uma partida de RPG: os(as) personagens e o(a) Mestre (ou Narrador(a)). Mestre é a pessoa que narra a partida, que faz a mediação entre as pessoas que jogam, o Sistema do Jogo e a história que está sendo contada. À pessoa que Mestra é destinado o desafio de estudar profundamente o enredo da partida e o Universo em que o jogo está acontecendo. Em nosso caso, a pessoa que Mestra não só deve saber o enredo e o Universo do jogo, mas talvez principalmente o conteúdo de Ensino a ser abordado, e portanto cabe ao(à) Docente este papel, ou a algum(a) estudante já treinado(a). Por fim, os(as) personagens, que são os(as) protagonistas da partida, são os papéis que cada jogador(a) terá que encenar durante o jogo. São os(as) personagens que agirão a partir do enredo narrado pelo(a) Mestre. O enredo inicial é contado pelo(a) Mestre, e a partir dele as personagens promovem Ações, que podem resultar em Acertos ou Erros, e assim o grupo todo vai desbravando o ambiente em questão.

\section{A Vingança de Newton}

Neste trabalho mostramos um jogo de RPG criado pelos autores, para estimular uma Metodologia de Ensino de Física Moderna por meio de um processo investigativo. O ferramental utilizado é a estratégia de Gamificação, por meio de um Role Playing Game (RPG) \cite{6,7,8,9,10,11,12,13}, com foco em Física Moderna.

\subsection{Enredo Principal}

No enredo inicial, o físico Isaac Newton está muito furioso, uma vez que vários cientistas começaram a questionar a validade da Teoria Newtoniana. Vários cientistas famosos da história da humanidade são levados para o Universo Paralelo em que Newton se encontra, onde as Leis Newtonianas não funcionam. Além disso, incluímos como personagem NPC (Non Player Character, um personagem que não joga, mas que pode servir para informações dentro do jogo) a \emph{Maçã Maluca}, como alívio cômico da Aventura.

Sendo assim, o trabalho dos estudantes, jovens cientistas, é de solucionar os desafios que forem surgindo ao longo do caminho. O primeiro passo para isso é definir qual personagem eles querem ser: afinal, em um jogo de RPG, o jogador interpreta um personagem de sua escolha. Os estudantes poderão escolher um dos diversos personagens dentre cientistas da história da física e matemática, disponibilizados em uma Biografia já fornecida pelos autores. Os arquivos relacionados à este trabalho, como Manual do jogador(a) e Biografias estão disponíveis por demanda, basta solicitar aos autores. O Manual do Mestre pode ser requisitado por e-mail, pedimos isto pois ele contém dados sobre as aventuras já produzidas.

\subsection{Atributos e Jogabilidade}

Dentro do Sistema que criamos, que é uma derivação do Sistema G\&G \cite{14}, cada personagem é descrito por 5 atributos: lógico-matemático, percepção, social, conceituação e experimentação; e tais atributos possuem valores que vão de 3 a 11, sendo 3 o menor peso (menor significância) para o atributo em questão e 11 o maior peso (maior significância). Essa relação surge a partir da rolagem de dois dados de 6 lados, onde o menor valor da soma dos dois dados a ser obtido com essa rolagem é 2 (1+1) e a maior rolagem obtida é 12 (6+6). Assim, escolhemos um valor acima do menor valor da soma e um valor abaixo do maior valor da soma para serem os limites inferiores e superiores dos atributos. A soma de todos os atributos de todos os personagens é sempre 33, como forma de garantir que personagens estão balanceados no jogo. A relação de cada personagem e seus respectivos atributos pode ser vista na tabela abaixo:

\begin{table}[h]
\centering
\begin{tabular}{|l|c|c|c|c|c|}
\hline
\textbf{Personagem} & \textbf{Lógico-matemático} & \textbf{Conceituação} & \textbf{Experimentação} & \textbf{Percepção} & \textbf{Social} \\
\hline
César Lattes       & 7  & 8  & 10 & 5  & 3 \\\hline
Emmy Noether       & 10 & 8  & 4  & 5  & 6 \\\hline
Mileva Maric       & 10 & 8  & 5  & 6  & 4 \\\hline
Erwin Schrodinger  & 10 & 8  & 4  & 5  & 6 \\\hline
Albert Einstein    & 8  & 10 & 3  & 5  & 7 \\\hline
Carol Nemes        & 9  & 10 & 3  & 7  & 4 \\\hline
Marie Curie        & 6  & 8  & 10 & 5  & 4 \\\hline
Richard Feynman    & 8  & 10 & 3  & 5  & 7 \\
\hline
Niels Bohr    & 7  & 10 & 8  & 5  & 3 \\
\hline
Ada Lovelace    & 10  & 6 & 8  & 5  & 4 \\
\hline
\end{tabular}
\caption{Relação personagens da aventura “A Vingança de Newton” e seus respectivos atributos.}
\end{table}

A jogabilidade é realizada da seguinte forma: para cada Ação um(a) personagem joga dois dados de 6 lados (dependendo da dinâmica da partida, quanto mais dados se jogar maior a chance de Acerto). A pessoa que Mestra deve analisar \emph{qual Atributo} usar em cada Ação e selecionado, a depender de \emph{que Ação} é executada. Podemos ter 3 saídas para a rolagem de dados: a) \textbf{Acerto simples}: personagem com Atributo selecionado que tira um valor nos dados MENOR que o Atributo na rolagem de dados; b) \textbf{Erro}: personagem com Atributo selecionado que tira um valor nos dados MAIOR que o Atributo na rolagem de dados; c) \textbf{Acerto crítico}: quando a pessoa tira na rolagem do dado exatamente o valor do Atributo selecionado.

Para cada \textbf{Acerto simples} a pessoa que Mestra o jogo fornece uma informação simples sobre o desafio, baseado no que o(a) personagem tentou agir na cena. Para cada \textbf{Erro}, há uma punição para o(a) personagem. E para cada \textbf{Acerto crítico} uma informação muito importante é dada para a pessoa. Em nossas partidas, estipulamos 3 pontos de vida para cada sala, sendo que cada punição gerava a perda de 1 ponto de vida (os pontos de vida eram ``resetados'' ao passar para a próxima sala).

\subsection{A Jogabilidade do sistema em Prática}

Seja o seguinte exemplo: se a personagem for \emph{Erwin Schrodinger} e seu jogador quis realizar a seguinte Ação: tentar observar se ao incidir Luz Violeta numa placa de metal um elétron ``pula'' da placa, criando uma certa corrente elétrica. O Atributo escolhido pelo Mestre pode ser, por exemplo, Experimentação, que neste caso vale 4 (ver Tabela 1). A pessoa então joga dois dados de 6 lados:
\begin{itemize}
\item[(i)] Se o resultado do dado for MENOR que 4 $\rightarrow$ Acerto simples: a pessoa que Mestra pode dar a Informação: ``Você notou que há uma corrente elétrica se formando. Mas não conseguiu medir o valor da mesma.''
\item[(ii)] Se o resultado for MAIOR que 4 $\rightarrow$ Erro, então punição: ``Você tentou jogar a Luz Violeta na placa, mas acertou a parede. Perdeu um ponto de vida.'' \item[(iii)] Se o resultado for igual a 4 $\rightarrow$ Acerto crítico: ``Além de notar que há uma corrente, você mediu exatamente a corrente que passa no circuito, e notou que se mudarmos a Voltagem a corrente diminui.''
\end{itemize}

\section{Um exemplo de Aventura: O Efeito \emph{FotoNewtoniano}}

Uma aventura completa já está totalmente estruturada, e visa a compreensão do Efeito Fotoelétrico (chamado na partida de \emph{Efeito FotoNewtoniano}). Ela é dividida em algumas etapas: uma introdutória, onde é narrado o enredo da história e como as etapas funcionarão; as Salas, onde estarão os desafios a serem cumpridos, que é, de fato, o jogo acontecendo e, por fim, a Sala Final, onde serão realizadas as considerações por parte do(a) Mestre(a) acerca da partida de RPG (e onde Newton se redime com seus colegas cientistas, chegando à conclusão que a Ciência progride a partir do trabalho em conjunto das pessoas).

O intuito da aventura já elaborada é que as pessoas de cada equipe (grupo) trabalhem em conjunto para solucionar os problemas de cada sala, sendo que o grupo deve resolver o problema daquele local para seguir adiante (Salas em sequência). Os conhecimentos adquiridos em cada uma das salas serão de suma importância para o entendimento dos desafios posteriores.

Nas seções abaixo mostramos como as salas estão atualmente estruturadas, para trabalhar o tema específico de cada sala (lembrando que a aventura do \emph{Efeito FotoNewtoniano} possui 4 salas sequenciais, sendo a última o encerramento da partida).

\subsection{Sala 1 - A dualidade}

\begin{figure}[h]
 \centering
 \fbox{\includegraphics[width=0.8\textwidth]{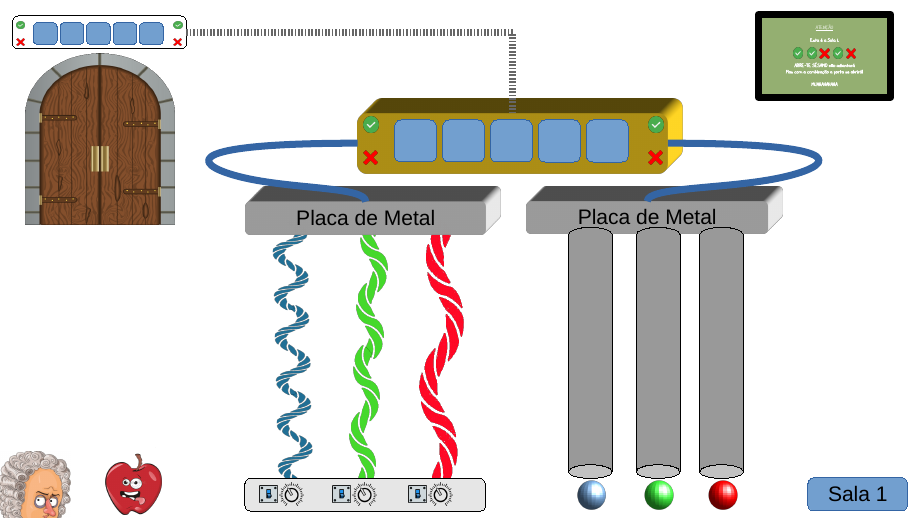}}
 \caption{Primeira sala da aventura do Efeito \emph{FotoNewtoniano}. Na imagem é possível ver uma porta (que está trancada inicialmente) com um sistema de chave em cima. Este sistema está conectado a um aparelho dourado que possui a mesma aparência: 5 quadrados azuis, com símbolos de ``correto'' e ``errado''. Este aparelho dourado está ligado, por sua vez, a duas placas de metal: uma placa possui 3 cordas, sendo uma azul, uma verde e uma vermelha, além disso cada corda está conectada a um botão \emph{on/off} e a um \emph{knob}/potenciômetro; a outra placa possui 3 canos, e na entrada de cada cano há uma bola, cada bola possui uma cor, sendo azul, verde e vermelho. Newton e sua amiga Maçã Maluca estão observando no canto da sala. Além disso, é possível ver um quadro de giz com algo escrito num canto da sala, oposto à porta.}
 \label{Fig1}
\end{figure}

Na primeira sala os jogadores surgem numa sala com os dispositivos mostrados na Figura \ref{Fig1}, sendo que a ideia desta primeira sala é servir como uma espécie de tutorial para jogadores, e também para já podermos trabalhar um pouco de conceitos como dualidade onda-partícula. A proposta é que estudantes/jogadores(as) vasculhem a sala e consigam solucionar o problema (que consta no quadro de giz, ver Figura \ref{Fig1} da primeira sala): descobrir qual a sequência vai abrir a porta. Em nossa proposta, a sequência é pré-definida, e consta no quadro de giz (ver Figura \ref{Fig1}), e está relacionada à cor da corda/bolinha, sendo que não importa se jogadores(as) usem cordas, ou bolinhas coloridas, nem a ``intensidade'' de vibração da corda, mas somente se usam um objeto (corda ou bolinha) com uma ``cor'' suficiente que vai acionar o aparelho. A ideia é que, ao final, quem estiver trabalhando com este problema instrua aos(às) jogadores(as) que a ``cor'' do objeto está relacionada à Energia do mesmo, para acionar o aparelho e abrir a porta.

\subsection{Sala 2 - O Efeito \emph{FotoNewtoniano} - A Intensidade da Luz}

\begin{figure}
 \centering
 \fbox{\includegraphics[width=0.8\textwidth]{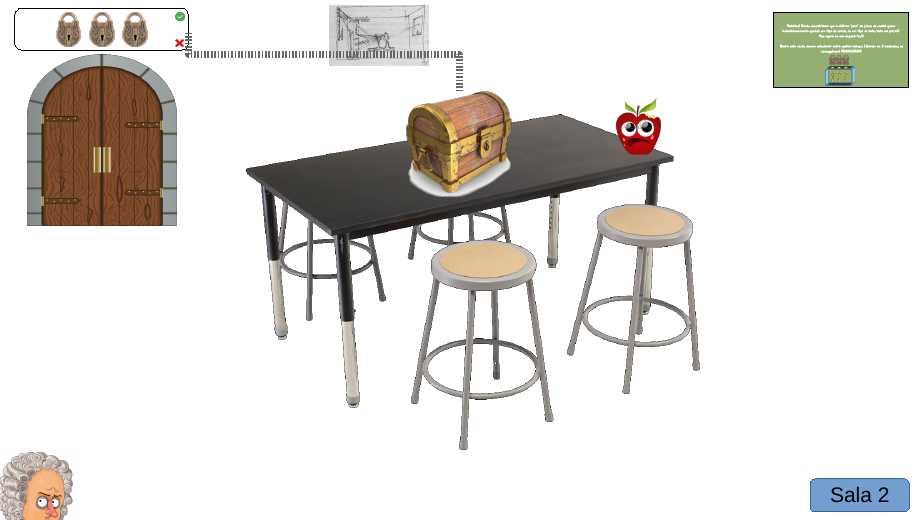}}
 \caption{Segunda sala da aventura do Efeito \emph{FotoNewtoniano}. Na imagem podemos ver outra porta, também inicialmente trancada. Acima da porta, vê-se três cadeados, acoplados com um sistema de símbolos ``correto'' e ``errado''. Os cadeados estão ligados a um baú, que está sobre uma mesa, no meio da sala. A Maçã Maluca está mordida, com cara de assustada, em cima da mesa. Numa das paredes pode-se ver um quadro com desenhos, e no canto oposto à porta há outro quadro de giz, com dizeres diferentes. Newton continua a observar num canto.}
 \label{Fig2}
\end{figure}

\begin{figure}
 \centering
 \fbox{\includegraphics[width=0.8\textwidth]{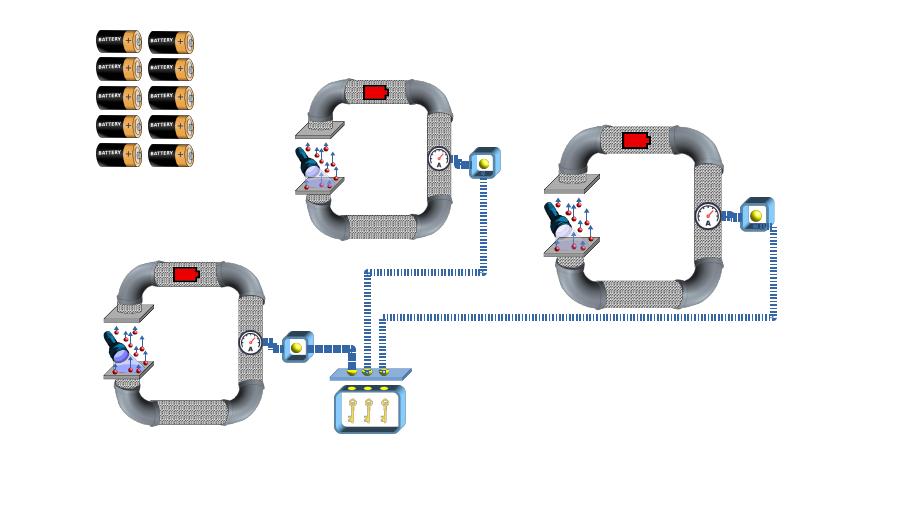}}
 \caption{Nesta imagem mostramos detalhes do que há dentro do baú da segunda sala: três aparelhos idênticos, aparentemente todos formados por canos. Em cada aparelho há uma lanterna, emitindo Luz da mesma cor porém de brilhos diferentes (note que a informação sobre o brilho NÃO é dada inicialmente aos jogadores, somente se perguntado), que incidem numa parte do aparelho. Onde a Luz incide, algumas ``bolinhas'' saltam da placa de incidência. Cada aparelho possui um medidor, que por sua vez está conectado a uma espécie de compartimento, que possui uma bolinha. Cada compartimento que possui bolinha está conectado a outro compartimento, onde é possível ver três chaves. Ao lado de cada aparelho podemos ver dez baterias, todas possuem a mesma tensão/voltagem.}
 \label{Fig3}
\end{figure}

Ao passarem para a segunda sala, o grupo encontra-se no ambiente que mostramos na Figura \ref{Fig2}. Na sala agora podemos ver uma mesa, com um baú. As instruções também são dadas no quadro de giz, e o grupo deve novamente solucionar o \emph{puzzle} de modo a abrir a porta. Dentro do baú o grupo encontra três espécies de ``circuitos'' ou aparelhos, que decidimos ilustrar como semelhantes a encanamentos, ver Figura \ref{Fig3}. Em cada aparelho há uma lanterna, emitindo Luz da mesma cor porém de brilhos diferentes (note que a informação sobre o brilho NÃO é dada inicialmente aos jogadores, somente se perguntado), que incidem numa parte do aparelho. Onde a Luz incide, algumas ``bolinhas'' saltam da placa de incidência. Cada aparelho possui um medidor, que por sua vez está conectado a uma espécie de compartimento, que possui uma bolinha. Cada compartimento que possui bolinha está conectado a outro compartimento, onde é possível ver três chaves. Ao lado de cada aparelho podemos ver dez baterias, todas possuem a mesma tensão/voltagem.

Nosso objetivo nesta sala é: que cada grupo consiga notar que será necessária a mesma quantidade de baterias para ``zerar'' o fluxo de ``bolinhas'' que saltam da placa, dado que a lanterna emite Luz da mesma cor, \emph{independente} do brilho da lanterna, que é fixo em cada aparelho porém diferente de um aparelho para outro. Nosso grupo de pesquisa estipulou o número de baterias como 7(sete) para a lanterna com Luz violeta, de forma arbitrária (de fato, o número sete é devido à terceira sala, onde a tarefa é identificar que o número de baterias nos circuitos varia dependendo da cor da lanterna).

\subsection{Sala 3 - O Efeito \emph{FotoNewtoniano} - A ``cor'' da Luz}

\begin{figure}
 \centering
 \fbox{\includegraphics[width=0.8\textwidth]{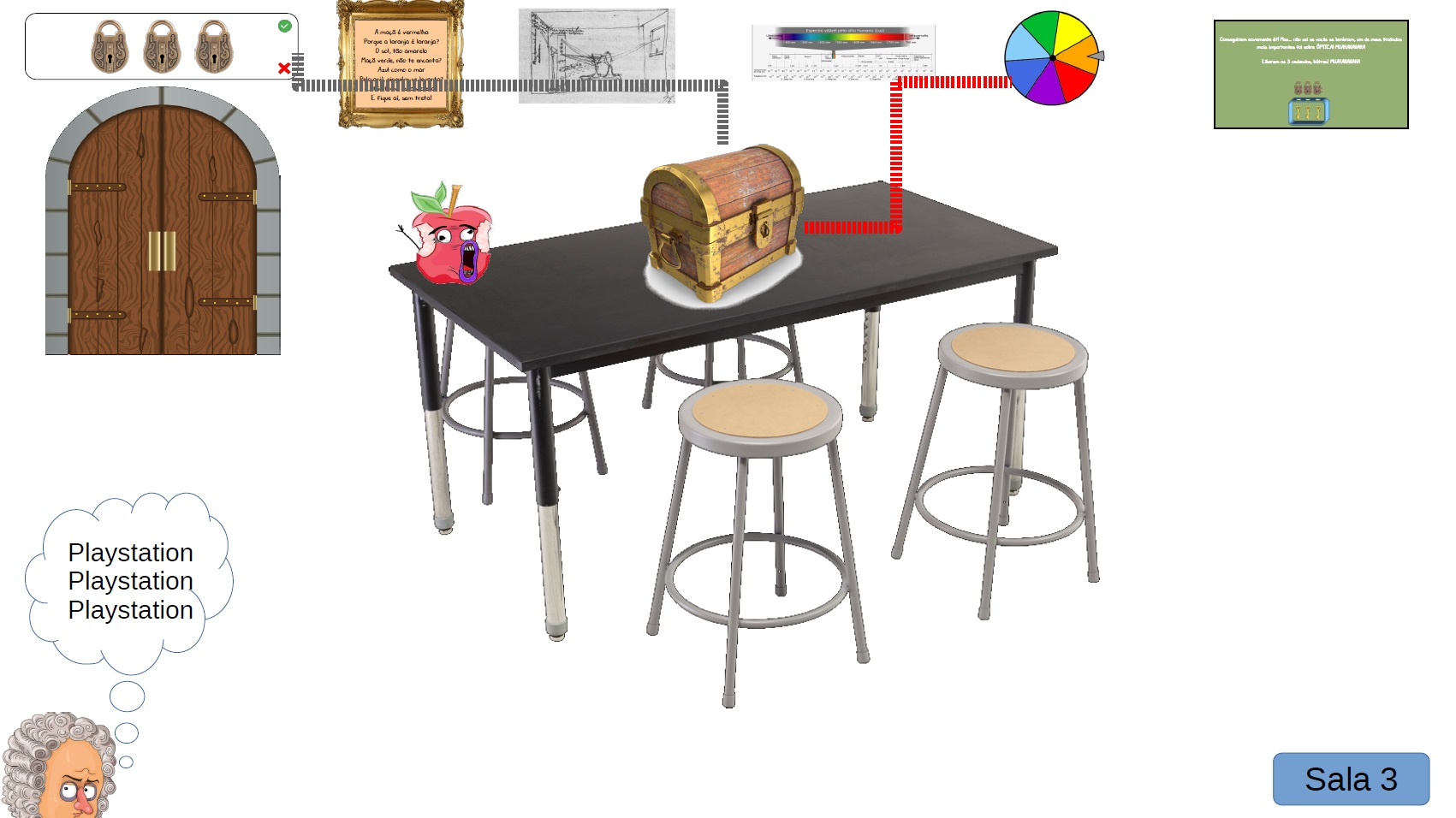}}
 \caption{Terceira sala da aventura do Efeito \emph{FotoNewtoniano}. Na imagem podemos ver uma porta, também inicialmente trancada. Acima da porta, vê-se três cadeados, acoplados com um sistema de símbolos ``correto'' e ``errado''. Os cadeados estão ligados a um baú, que está sobre uma mesa, no meio da sala. Uma roleta com 7 cores está presa na parede e também está conectada ao baú. A Maçã Maluca está com duas mordidas, com cara de assustada e com a boca roxa, em cima da mesa (a Maçã Maluca de fato dá dicas sobre cada sala). Numa das paredes pode-se ver um quadro com desenhos como na segunda sala, mas também vemos um quadro antigo com uma poesia, e um cartaz com o espectro eletromagnético. No canto oposto à porta há outro quadro de giz, com dizeres diferentes. Newton continua a observar num canto e pensa ``Playstation, Playstation, Playstation''.}
 \label{Fig4}
\end{figure}

\begin{figure}
 \centering
 \fbox{\includegraphics[width=0.8\textwidth]{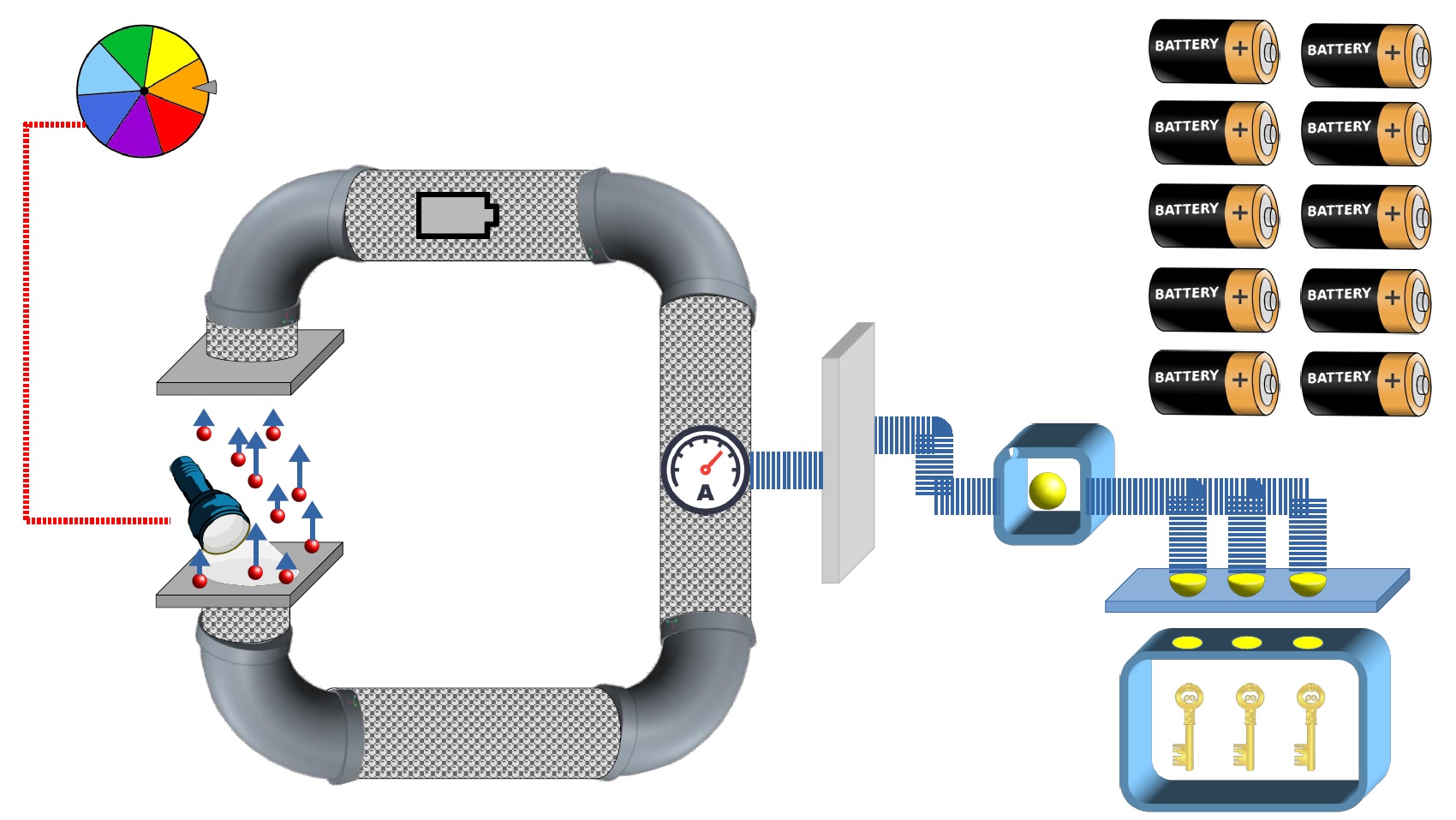}}
 \caption{Nesta imagem mostramos detalhes do que há dentro do baú da terceira sala: um aparelho, também formado por canos. Há uma lanterna conectada ao aparelho, emitindo inicialmente Luz branca, que incide numa parte do aparelho. Onde a Luz incide, algumas ``bolinhas'' saltam da placa de incidência. Cada aparelho possui um medidor, que por sua vez está conectado a uma espécie de compartimento, que possui uma bolinha. O compartimento que possui bolinha está conectado a uma placa, inicialmente cinza, e a placa está ligada a outro compartimento, onde é possível ver três chaves e três espaços para a bolinha. Ao lado do aparelho podemos ver dez baterias, todas possuem a mesma tensão/voltagem. Ligado ao aparelho podemos ver uma roleta com sete cores, a roleta está conectada à lanterna.}
 \label{Fig5}
\end{figure}

Agora na terceira sala, o grupo se depara com um lugar como na Figura \ref{Fig4}. Continuamos vendo a mesa, também com um baú. As instruções também são dadas no quadro de giz, e a ideia é novamente que o grupo abra a porta. Nesta sala, além do mesmo quadro com desenho que a sala anterior possuía, há um quadro antigo contendo uma poesia\footnote{\emph{``A maçã é vermelha \\
Porque a laranja é laranja? \\
O sol, tão amarelo \\
Maçã verde, não te encanta? \\
Azul como o mar \\
Pelo anil, ninguém se levanta? \\
Então tome essa violeta \\
E fique aí, sem treta!''}}, um cartaz com o espectro eletromagnético enfatizando a região visível, e uma roleta com 7 cores. A roleta está também conectada ao baú. Dentro do baú, agora, o grupo encontra apenas um tipo de ``circuito'' ou aparelho, ver Figura \ref{Fig5}. No aparelho vemos uma lanterna, emitindo Luz \emph{branca} inicialmente, que incide numa parte do aparelho. Onde a Luz incide, algumas ``bolinhas'' saltam da placa de incidência novamente. Cada aparelho possui o mesmo medidor, mas dessa vez o medidor está conectado a uma placa cinza, e esta placa está conectada ao compartimento contendo uma bolinha. Este compartimento está conectado a outro, onde é possível ver também três chaves e três entradas para bolinhas. Ao lado do aparelho podemos ver novamente dez baterias, todas possuem a mesma tensão/voltagem.

Um detalhe importante nesta sala é que a roleta está conectada ao circuito dentro do baú. A cada rodada o grupo deverá lançar um dado, e dependendo do resultado podem receber informações sobre a roleta, e escolher uma das cores:
\begin{itemize}
\item acerto simples no dado: personagem escolhe a cor e a cor é a que vai acender na lanterna.
\item acerto crítico no dado: o mesmo do acerto simples, porém o personagem pode receber alguma dica a mais ou ter um insight.
\item erro: a cor que a pessoa quiser que acenda na lanterna não vai acender, acontece alguma fatalidade como a luz não acender.
\end{itemize}

Em nossa experiência realizando jogos dentro desta abordagem, pudemos observar que também há a opção da roleta funcionar rolando um dado de 7 lados (no Google basta digitar ``1d7'' que ele joga este dado automaticamente), e o resultado sendo: 1 $\rightarrow$ vermelho; 7 $\rightarrow$ violeta; e seguindo as cores na ordem. Caso a roleta funcione assim, a partida fica mais rápida e dinâmica.

Apenas uma certa quantidade de pilhas, colocadas da maneira correta, zera o marcador, e libera a bolinha, e consequentemente uma chave (outra bolinha reaparece após sair do compartimento). Nosso grupo estipulou o seguinte \emph{gabarito} para o uso das baterias com as cores: A quantidade correta de pilhas para zerar o marcador e liberar a bolinha, para cada cor de Luz que incide na placa, é:
\begin{itemize}
\item {\colorbox{red}{Vermelho}}: 1 pilha
\item {\colorbox{orange}{Laranja}}: 2 pilhas
\item {\colorbox{yellow}{Amarelo}}: 3 pilhas
\item {\colorbox{green}{Verde}}: 4 pilhas
\item {\colorbox{lightblue}{\color{white}{Azul}}}: 5 pilhas
\item {\colorbox{indigo}{\color{white}{Anil}}}: 6 pilhas
\item {\colorbox{violet}{\color{white}{Violeta}}}: 7 pilhas
\end{itemize}

Mas note que zerar o marcador não significará que a bolinha vai cair. Há uma placa especial entre o marcador e a bolinha. As cores pré estabelecidas (pelo grupo, isto pode ser alterado de acordo com a proposta de quem mestra) para que cada bolinha retire uma chave para o grupo são:
\begin{itemize}
\item {\colorbox{red}{Vermelho}}: 1 pilha
\item {\colorbox{green}{Verde}}: 4 pilhas
\item {\colorbox{violet}{\color{white}{Violeta}}}: 7 pilhas
\end{itemize} Ou seja: se o grupo escolher a cor amarela para a lanterna, e colocar 3 pilhas, o marcador vai zerar, essa informação será dada, porém a bolinha não irá se mexer, pois a placa continuará cinza.

Nosso objetivo nesta sala é: observar que a Frequência da Luz incidida \emph{altera} a energia mínima necessária para que as bolinhas ``saltem'' da placa (independente da Intensidade). Com isto, espera-se que o grupo chegue a conclusão que dependendo da \emph{cor} da Luz necessita-se de uma quantidade maior de Energia (dada pelas baterias) para zerar o marcador, isto é, para zerar o fluxo de elétrons no circuito.

\section{Sobre a Pesquisa em Ensino relacionada ao jogo  \emph{A Vingança de Newton}}

O objetivo principal de nosso jogo \emph{A Vingança de Newton} é duplo: (i) a disponibilização do jogo \emph{A Vingança de Newton}, com enredos prontos, de modo a poder ser utilizado livremente pela comunidade de Ensino de Ciências no Brasil; (ii) a análise dos dados obtidos, verificando a hipótese da Gamificação poder ser uma metodologia que contribui para o aprendizado específico de Física Moderna (note que a Mecânica de jogo que estamos propondo pode ser facilmente direcionada para qualquer área de Ensino).

Temos duas questões de pesquisa:
\begin{enumerate}
 \item há ``ganho de aprendizado''\footnote{Os(as) autores(as) entendem que o conceito de ``ganho de aprendizado'' é controverso, porém, pragmaticamente, decidimos por escolher a metodologia de pré e pós-teste, de modo a ``mensurar'' este ganho com, por exemplo, o ganho de Hake\cite{16}.} ao usar nosso jogo com um grupo de estudantes?
 \item estudantes \emph{percebem} uma metodologia que usa jogos como um processo de aprendizado? Esta pergunta surge no âmbito da referência \cite{percepcao}, onde os autores notaram que, apesar de estudantes que participaram de aulas com metodologias alternativas de ensino terem tirado notas maiores que estudantes em turmas ``tradicionais'', o primeiro grupo não teve a \emph{percepção} que estavam aprendendo nas aulas alternativas, o que levanta a hipótese de podermos estar condicionados a entender uma aula como sendo somente uma aula ``cuspe e giz''.
\end{enumerate}

A Metodologia de Pesquisa proposta visa coletar os seguintes dados:
\begin{itemize}
 \item resultados de pré e pós-testes (realizados via \emph{Google Forms}): os pré e pós-testes devem ser aplicados logo antes e depois de cada partida. Usamos uma metodologia onde os pré e pós-testes são compostos por perguntas iguais, com base em escala Likert;
 \item percepção de jogadores(as) com relação ao impacto do jogo de RPG como um mecanismo de Ensino.
 \item entrevistas: para análise de entrevistas, utilizaremos a sistemática de Análise de Conteúdo como descrito em Bardin \cite{15}.
\end{itemize}

Propomos a obtenção de dados quantitativos (por meio de questionários) e qualitativos (por meio de entrevistas). Por completeza, indicamos que as operações mínimas para se realizar uma pesquisa por meio de análise de conteúdo pode ser sintetizada como \cite{15}: (A) Delimitação dos objetivos e de quadro de referência teórico: em nosso caso, os objetivos principais são: utilizar um Método de Ensino Centrado no(a) Estudante baseado em Gamificação com o uso de RPG, focando num conteúdo de Física Moderna \cite{6,7,8,9,10,11,12,13}; (B) Constituição do Corpus: o material que coletaremos será composto por entrevistas com participantes das Mesas, questionários online sobre o Jogo, pré e pós-teste e suas respectivas notas/conceitos; (C) Definição de categorias e das unidades de análise: como citado em \cite{15}, as categorias, ou as classificações, para a análise do conteúdo podem ser definidas a priori ou a posteriori (ou mesmo uma mistura destes); (D) Tratamento dos resultados, inferência e interpretação: neste ponto analisaremos os dados quantitativos coletados, em especial os resultados de pré e pós-teste e dos questionários com escala Likert, por exemplo com utilização do Ganho de Hake \cite{16} e alfa de Cronbach \cite{17}.

\subsection{Pré e pós-testes}

Abaixo mostramos as perguntas que confeccionamos como pré e pós-teste para a aventura \emph{Efeito FotoNewtoniano}, sendo que aplicamos o mesmo questionário em ambas as situações:

\begin{enumerate}
 \item Marque o que você pensa sobre a seguinte afirmação: assim como os corpos são formados por átomos, a Luz é formada por pequeníssimas partículas de Luz. Na escala abaixo, marque sua resposta considerando: (1) Discordo totalmente; (2) Discordo em partes; (3) Nem concordo, nem discordo (indiferente); (4) Concordo em partes; (5) Concordo totalmente.
 \item Marque o que você pensa sobre a seguinte afirmação: assim como as ondas formadas quando jogamos uma pedrinha num lago, a Luz é uma onda que se propaga no espaço. Na escala abaixo, marque sua resposta considerando: (1) Discordo totalmente; (2) Discordo em partes; (3) Nem concordo, nem discordo (indiferente); (4) Concordo em partes; (5) Concordo totalmente.
 \item Marque o que você pensa sobre a seguinte afirmação: é possível gerar corrente elétrica se incidirmos Luz em determinados dispositivos. Na escala abaixo, marque sua resposta considerando: (1) Discordo totalmente; (2) Discordo em partes; (3) Nem concordo, nem discordo (indiferente); (4) Concordo em partes; (5) Concordo totalmente.
 \item Marque o que você pensa sobre a seguinte afirmação: quanto maior for o brilho de uma lâmpada, maior será a Energia que ela emite. Na escala abaixo, marque sua resposta considerando: (1) Discordo totalmente; (2) Discordo em partes; (3) Nem concordo, nem discordo (indiferente); (4) Concordo em partes; (5) Concordo totalmente.
 \item Marque o que você pensa sobre a seguinte afirmação: Luz Azul e Luz Vermelha se propagando no ar possuem a mesma velocidade. Na escala abaixo, marque sua resposta considerando: (1) Discordo totalmente; (2) Discordo em partes; (3) Nem concordo, nem discordo (indiferente); (4) Concordo em partes; (5) Concordo totalmente.
 \item Marque o que você pensa sobre a seguinte afirmação: você tem duas lanternas, uma emite Luz Azul, a outra Luz Vermelha, mas ambas possuem o mesmo brilho, portanto ambas emitem a mesma Energia. Na escala abaixo, marque sua resposta considerando: (1) Discordo totalmente; (2) Discordo em partes; (3) Nem concordo, nem discordo (indiferente); (4) Concordo em partes; (5) Concordo totalmente.
\end{enumerate}

\subsection{Percepção de aprendizado}

Abaixo mostramos as perguntas relacionadas à percepção de aprendizado dos(as) estudantes com relação ao jogo de RPG:

\begin{enumerate}
 \item Qual seu nível de conhecimento sobre física? Na escala abaixo, marque sua resposta considerando: (1) Muito ruim; (2) Ruim; (3) Nem ruim, nem bom (indiferente); (4) Bom; (5) Muito bom.
 \item O quanto você conhece sobre métodos de ensino não tradicionais? Na escala abaixo, marque sua resposta considerando: (1) Não conheço; (2) Conheço um pouco; (3) Conheço moderadamente; (4) Conheço muito; (5) Conheço extremamente.
 \item Qual seu nível de conhecimento sobre RPGs (Role Playing Games)? Na escala abaixo, marque sua resposta considerando: (1) Não conheço; (2) Conheço um pouco; (3) Conheço moderadamente; (4) Conheço muito; (5) Conheço extremamente.
 \item Qual seu nível de conhecimento sobre jogos educativos? Na escala abaixo, marque sua resposta considerando: (1) Não conheço; (2) Conheço um pouco; (3) Conheço moderadamente; (4) Conheço muito; (5) Conheço extremamente.
\item Qual seu nível de conhecimento sobre história da ciência? Na escala abaixo, marque sua resposta considerando: (1) Não conheço; (2) Conheço um pouco; (3) Conheço moderadamente; (4) Conheço muito; (5) Conheço extremamente.
\item Para você, quão útil é o uso de jogos (tabuleiro, RPG, video-game, cartas, etc.) para aprender uma matéria, como Física por exemplo? Na escala abaixo, marque sua resposta considerando: (1) Nem um pouco útil; (2) Um pouco útil; (3) Mais ou menos útil; (4) Muito útil; (5) Extremamente útil.
\item Comparando quando você estuda em grupo (quando você estuda com mais pessoas) com quando você estuda sozinho(a) (individualmente), qual você considera mais efetivo? Na escala abaixo, marque sua resposta considerando: (1) Totalmente individual; (2) Um pouco mais individual; (3) Nem só individual, nem só em grupo; (4) Um pouco mais em grupo; (5) Totalmente em grupo.
\item Qual o seu nível de interesse em Física? Na escala abaixo, marque sua resposta considerando: (1) Nenhum; (2) Pouco interesse; (3) Interesse moderado; (4) Muito interesse; (5) Interesse extremo.
\end{enumerate}

\section{Conclusão}

Em resumo, apresentamos o sistema de jogo desenvolvido por nosso grupo, e o tema base do jogo de RPG \emph{A Vingança de Newton}. Possuímos atualmente uma aventura completa, e questionários de avaliação para mensurar tanto ``ganho de aprendizado'' quanto a percepção de estudantes quanto à metodologia. Uma aventura quase finalizada também existe, para tratar de estrutura eletrônica básica. Acreditamos que nosso sistema é simples o suficiente para ser aplicado em diversas áreas, e deixamos o convite à comunidade que, caso queiram, entrem em contato para utilizar nosso jogo e também nossos questionários, de modo a obter dados com amostragem numerosa para serem devidamente analisados.

Os resultados apresentados neste estudo indicam que o uso de jogos de RPG como metodologia ativa de ensino pode ser uma ferramenta poderosa no ensino de Física Moderna. A implementação do jogo A Vingança de Newton pode ser eficaz, tanto no engajamento dos estudantes quanto na promoção de uma melhor compreensão de conceitos científicos, como o efeito fotoelétrico. Através de atividades investigativas e colaborativas, os estudantes podem ser levados a desenvolver habilidades cognitivas e sociais, alinhadas às teorias construtivistas de Piaget e Vygotsky.

Além disso, dados qualitativos podem ser obtidos por meio de entrevistas e questionários, e podem por sua vez reforçar o RPG como uma metodologia divertida e eficaz para o ensino de ciências. Dessa forma, concluímos que a gamificação, particularmente por meio de jogos de RPG, tem o potencial de transformar o ensino de física ao envolver os estudantes de forma mais ativa e participativa. Encorajamos novas pesquisas para explorar ainda mais o impacto de jogos educativos em outras áreas do conhecimento.

Enfatizamos que toda a documentação de nosso sistema de jogo e da aventura apresentada (Manual do Mestre, Manual do Jogador, Desenhos das salas e detalhes) podem ser solicitados por e-mail.

\section*{Comitê de Ética em Pesquisa com Seres Humanos}

Salientamos que dentre os meses de fevereiro a abril de 2022, foi realizada a submissão e correções solicitadas desta Metodologia no Comitê de Ética em Pesquisa com Seres Humanos, da Plataforma Brasil e a posterior aprovação do projeto. Assim sendo, a pesquisa e, consequentemente, o RPG como um todo, estão aptos para serem aplicados em escolas e universidades a nível médio e superior, respectivamente, bem como a aplicação dos questionários para Pesquisa em Ensino de Física (Certificado de Apresentação de Apreciação Ética - CAAE: 56933822.0.0000.5153).

\begin{acknowledgement}
Os autores agradecem as seguintes instituições pelo auxílio financeiro e logístico: CNPq (Conselho Nacional de Desenvolvimento Científico e Tecnológico), FAPEMIG (Fundação de Amparo à Pesquisa do Estado de Minas Gerais), FUNARBE (Fundação Arthur Bernardes) através de bolsa de Iniciação à Pesquisa em Ensino - FUNARBEN, INCT-IQ (Instituto Nacional de Ciência e Tecnologia de Informação Quântica).
 \end{acknowledgement}


\end{document}